\documentclass[structabstract]{aa}  
\usepackage{graphicx}
\usepackage{txfonts}
\usepackage{natbib}
\usepackage{hyperref}
\bibpunct{(}{)}{;}{a}{}{,}
\begin{document}
   \title{On the functional form of the metallicity-giant planet correlation\thanks{Tables 1 and 2 are only available in electronic form
at the CDS via anonymous ftp to cdsarc.u-strasbg.fr (130.79.128.5) or via http://cdsweb.u-strasbg.fr/cgi-bin/qcat?J/A+A/}}

   \subtitle{}

   \author{A. Mortier\inst{1,2},
           N.C. Santos\inst{1,2},
           S. Sousa\inst{1,3},
           G. Israelian\inst{3,4},
           M. Mayor\inst{5}
           \and
           S. Udry\inst{5}
          }

   \institute{Centro de Astrof\'{\i}sica, Universidade do Porto, Rua das Estrelas, 4150-762 Porto, Portugal\\
              \email{amortier@astro.up.pt}
	      \and
	      Departamento de F\'{\i}sica e Astronomia, Faculdade de Ci\^encias, Universidade do Porto, Portugal	      
	\and
	      Instituto de Astrof\'{\i}sica de Canarias, 38200 La Laguna, Tenerife, Spain 
	\and
	      Departamento de Astrof\'{\i}sica, Universidad de La Laguna, 38206 La Laguna, Tenerife, Spain 
	\and
	      Observatoire de Gen\`eve, Universit\'e de Gen\`eve, 51 Ch. des Maillettes, 1290 Sauverny, Switzerland
	      }

   \date{Received 09/11/2012; Accepted 28/01/2013}

 
  \abstract
   {It is generally accepted that the presence of a giant planet is strongly dependent on the stellar metallicity. A stellar mass dependence has also been investigated, but this dependence does not seem as strong as the metallicity dependence. Even for metallicity, however, the exact form of the correlation has not been established.}
   {In this paper, we test several scenarios for describing the frequency of giant planets as a function of its host parameters. We perform this test on two volume-limited samples (from CORALIE and HARPS).}
   {By using a Bayesian analysis, we quantitatively compared the different scenarios.}
   {We confirm that giant planet frequency is indeed a function of metallicity. However, there is no statistical difference between a constant or an exponential function for stars with subsolar metallicities contrary to what has been previously stated in the literature. The dependence on stellar mass could neither be confirmed nor be discarded.}
   {}

   \keywords{Planet-star interactions -- Stars: abundances -- Methods: statistical
               }

   \authorrunning{Mortier, A. et al.}
   \maketitle
%

\section{Introduction}

Since the discovery of the first extrasolar planet around a solar-like star in 1995 \citep[51 Peg b, ][]{Mayor95}, the search for extrasolar planetary systems has accelerated. As of today, more than 800 planets have been announced. Most of them were detected using the radial velocity technique. Although 800 is a relatively high number, the theory of planet formation and evolution is still being debated \citep{Pol96,May02,Mor09}. The situation is particularly difficult for giant planet formation. Currently, there are two proposed models: core accretion \citep[e.g.][]{Pol96,Rice03,Ali04}, where gas from the protoplanetary disk is accreted around a previously formed rocky/icy core; and the disk instability model \citep[e.g.][]{Boss97,May02}, where a planet is formed because of a direct gravitational instability in the protoplanetary disk, in the same way as stars form from interstellar clouds. A helpful overview of both models is given by \citet{Mat07}.

This problem may be solved with the help of planet-host stars. Observational and theoretical evidence shows that the presence of a planet seems to depend on several stellar properties, such as mass and metallicity \citep{Udry07}. Concerning metallicity, it has been well-established that more metal-rich stars have a higher probability of harboring a giant planet than their lower metallicity counterparts \citep{Gon97,San01,San04,Fis05,Udry07,Sou11b,Me12}. The occurrence rate even increases dramatically with increasing metallicity. Current numbers, based on the CORALIE and HARPS samples, suggest that around $25\%$ of the stars with twice the metal content of our Sun are orbited by a giant planet. This number decreases to $\sim5\%$ for solar-metallicity objects \citep{Sou11b,Mayor11}. A similar trend has also been found by previous results \citep[e.g.][]{San04,Fis05,John10}. Curiously, no such trend is observed for the lower mass planets \citep{Udry06,Sou08,Mayor11}. The Neptune-mass planets found so far seem to have a rather flat metallicity distribution \citep{Sou08,Sou11b,Mayor11}.

While this exponential trend is very clear for solar and supersolar metallicities, the situation for lower (subsolar) metallicities is still uncertain. \citet{San04} and \citet{Udry07} suggest that a constant frequency may be a better fit for subsolar metallicities than a continuous exponential. They used the sample from \citet{Fis05} as well as the CORALIE sample. Their suggestion has been discarded by \citet{John10}, based on a Bayesian analysis on a large SPOCS sample. Recently, it has also been suggested that there may be a lower limit below which no giant planets can be formed anymore \citep{Me12}. The issue of giant planet frequency dependence on stellar properties is still not resolved.

The observed metallicity correlation in the metal-rich region favors the core-accretion model for the formation of giant planets \citep{Ida04,Udry07,Mor12} because the higher the grain content of the disk, the easier it is to build the cores that will later accrete gas. According to the disk-instability model, however, the presence of giant planets would not be strongly dependent on stellar metallicity \citep{Boss02}. It is thus important that we fully understand what happens around stars with different metallicities and masses since it can provide clues to the processes of planet formation and evolution. Different formation mechanisms may play a role around stars with different metallicities. 

In the light of this ongoing debate, we analyze two volume-limited samples in this paper to find out how the situation looks in this low-metallicity end. In Section \ref{Dat}, an overview is given of the samples and their data. Section \ref{Bay} reports on the analysis of these samples. Finally, conclusions are made in Section \ref{Con}, together with a discussion.


\section{The samples}\label{Dat}

To test the giant planet frequency dependence on the stellar properties, we used two volume-limited samples. In each sample, there are several planet hosts. Since this paper is about giant-planet frequency, we considered planet hosts as only those stars with at least one planet with a mass between $0.1$ and $25$ Jupiter mass. These planets have periods between 5 and 5000 days. \citet{Mayor11} calculated the detection limits on these samples. They show that the maximum orbital period out to which a 0.1 Jupiter-mass planet could be detected lies around 1000 - 2000 days.

\subsection{Coralie sample}

The CORALIE sample is a large one of 1216 stars. It was constructed from the larger CORALIE sample that consisted of about 1650 stars \citep{Udry00}. A color cut-off was then made. Stars with a $B-V > 1.2$ were discarded from the sample. This leaves us with a sample of about $\sim 1250$ stars, as used in several previous works \citep[e.g. ][]{San04,Sou11b}.

Since high-resolution spectra were not available for all of these stars, metallicities were derived using the cross correlation function (CCF) calibration.
From the Hipparcos catalog \citep{VanL07}, the color $B-V$ was used to derive temperatures with the calibration formula, reported in \citet{Sou08}. Stellar masses were estimated following the same procedure as in \citet{Sou11a} where stellar evolutionary models from the Padova group were applied using their web interface. This could not be done for 37 stars, leaving us with a final sample of 1216 stars. The parameters can be found in Table \ref{TabCoralie}. In the bottom panel of Fig. \ref{FigCov}, the data is presented in a metallicity-mass diagram. Metallicity and mass for the stars in this sample have typical error bars of 0.07\,dex and 0.1\,M$_{\odot}$, respectively.

\begin{table}
\caption{Stellar parameters for the stars in the CORALIE sample. The complete table is provided in electronic form only. }
\label{TabCoralie}     
\centering                       
\begin{tabular}{lllll}
\hline\hline 
Star & $[Fe/H]$ & $M_{\ast}$ & $T_{eff}$ & planet?\\
 &  & $(M_{\odot})$ & $(K)$ & \\                
\hline
HD967 & -0.68 & 0.8 & 5557.82 & \\
HD1108 & 0.04 & 0.94 & 5656.61 & \\
HD1237 & 0.07 & 0.92 & 5506.26 & yes\\
HD1320 & -0.27 & 0.89 & 5688.25 & \\
HD1388 & -0.01 & 1.05 & 5967.46 & \\
HD1461 & 0.19 & 1.04 & 5780.9 & yes\\
... & ... & ... & ... & ... \\
\hline                       
\end{tabular}
\end{table}

\subsection{Harps sample}\label{Harps}

Our second sample was constructed as part of a HARPS GTO program that aims to detect and obtain accurate orbital elements of Jupiter-mass planets in a well-defined volume of the solar neighborhood \citep[out to 57.5 pc from the Sun - ][]{Loc10}. It can be seen as an extension to the CORALIE sample.

\begin{figure}[t!]
\begin{center}
\includegraphics[width=6.7cm,angle=270]{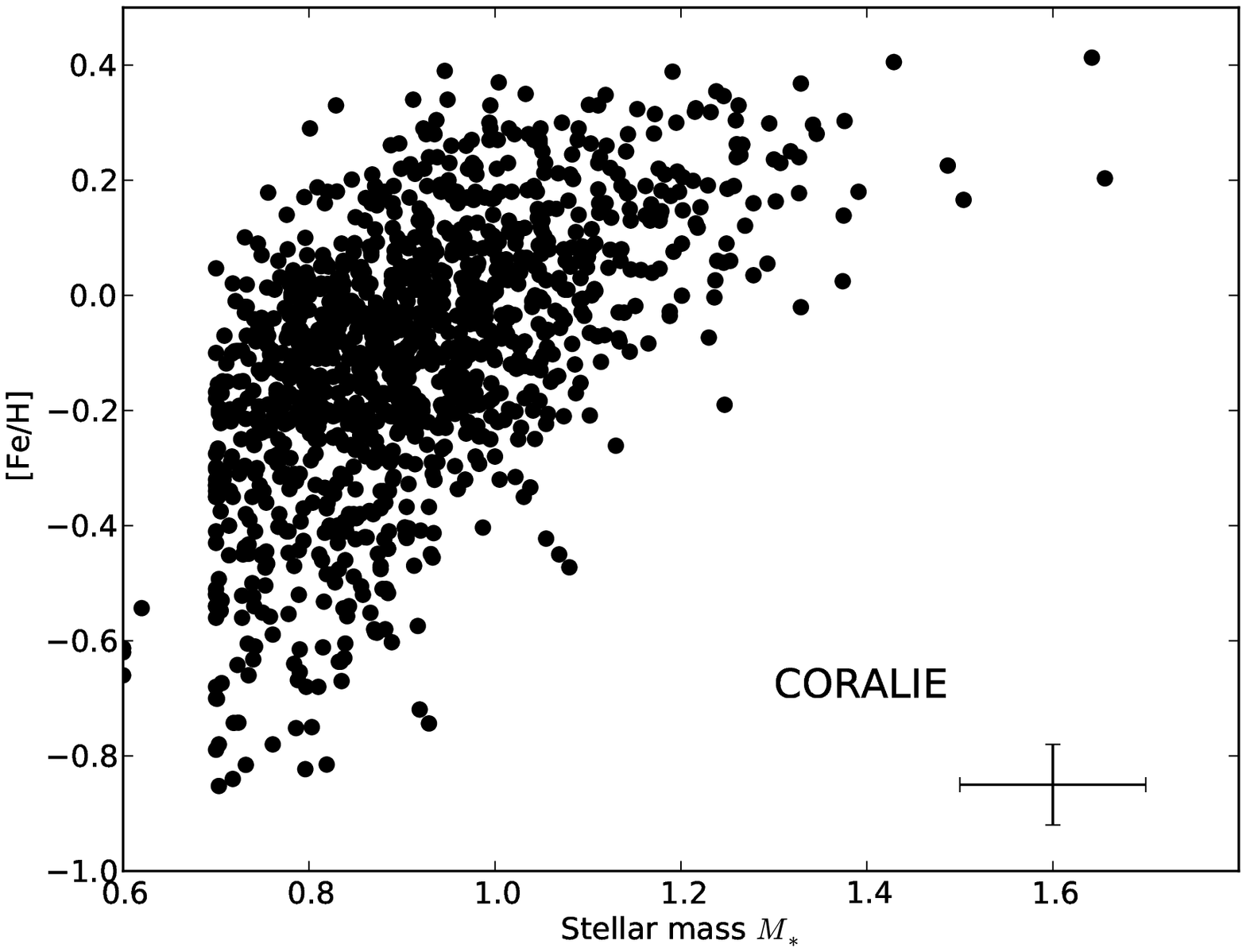}\\
\includegraphics[width=6.7cm,angle=270]{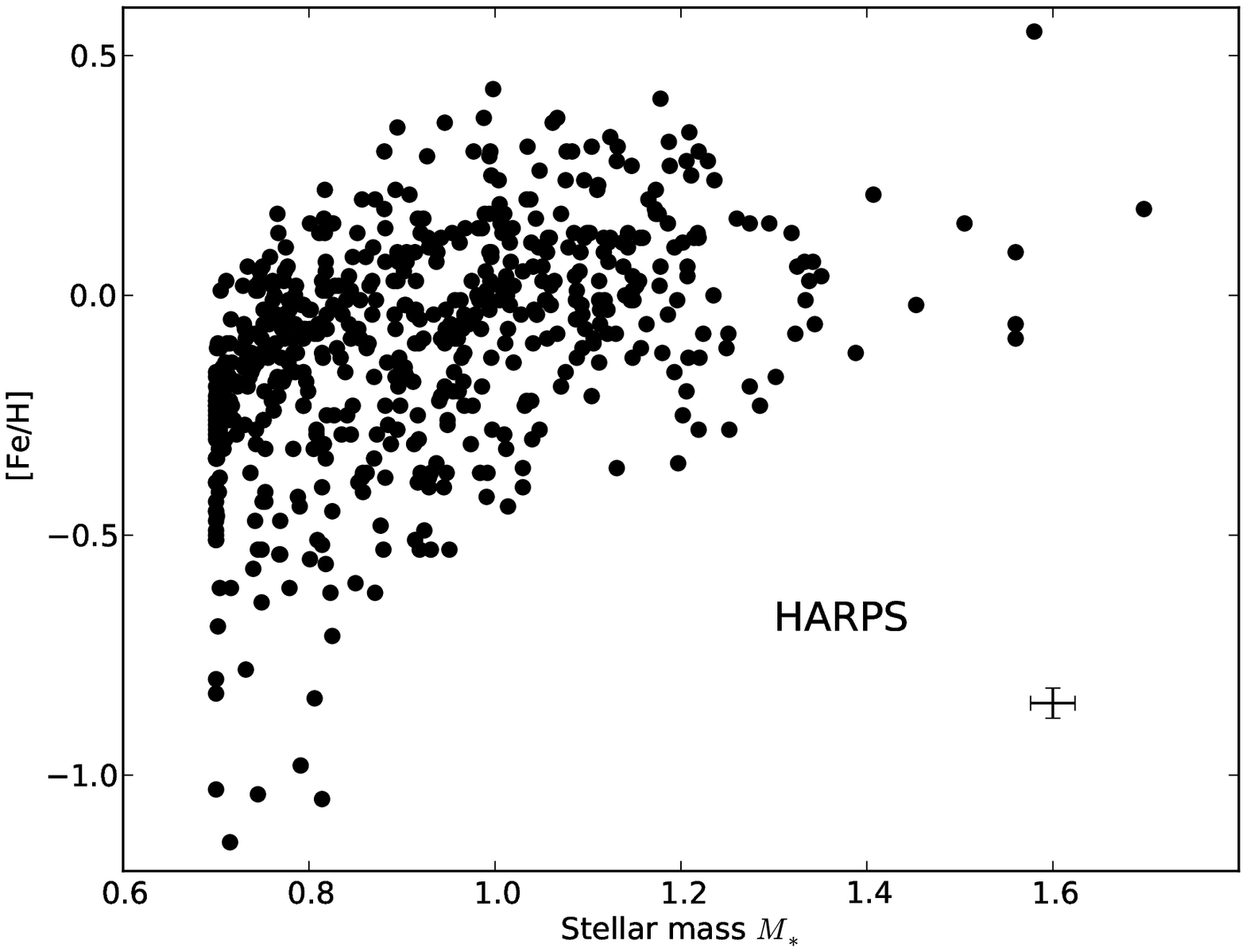}
\caption{Mass - metallicity coverage of the CORALIE (top panel) and the HARPS (bottom panel) sample.}
\label{FigCov}
\end{center}
\end{figure}

High-resolution spectra were taken with the HARPS spectrograph \citep{Mayor03}. In total, 582 stars had good enough spectra to spectroscopically derive accurate stellar parameters. The metallicities for these stars were all derived with the method described in \citet{San04}. The equivalent widths were automatically measured with the ARES code \citep[Automatic Routine for line Equivalent widths in stellar Spectra - ][]{Sou07}. \citet{Sou11b} compared metallicities, derived with this method and the CCf correlation, for a subsample of the CORALIE sample. The values are consistent with a mean difference of 0.01\,dex and a dispersion of 0.07\,dex. Stellar masses were estimated following the same procedure as in \citet{Sou11a}, where stellar evolutionary models from the Padova group were applied using their web interface. All parameters have been previously reported in \citet{Sou11b} and can also be found in Table \ref{TabHarps}. In the top panel of Fig. \ref{FigCov}, the data is presented in a metallicity-mass diagram. Metallicity and mass for the stars in this sample have typical error bars of 0.03\,dex and 0.025\,M$_{\odot}$, respectively.

\begin{table}
\caption{Stellar parameters for the stars in the HARPS sample. The complete table is provided in electronic form only. }
\label{TabHarps}     
\centering                       
\begin{tabular}{llll}
\hline\hline 
Star & $[Fe/H]$ & $M_{\ast}$ & planet?\\
 &  & $(M_{\odot})$ & \\                
\hline
HD5388 & -0.28 & 1.22 & \\
HD6718 & -0.07 & 0.95 & yes\\
HD8038 & 0.15 & 1.01 & \\
HD8535 & 0.04 & 1.15 & yes\\
HD8930 & -0.23 & 0.9 & \\
HD8985 & -0.01 & 1.2 & \\
... & ... & ... & ... \\
\hline                       
\end{tabular}
\end{table}

\section{Bayesian analysis}\label{Bay}

The analysis performed in this paper is based on the methods described in \citet{Kass95} and \citet{John10}. It uses a Bayesian inference to find the best relationship between stellar mass and metallicity and the presence of a giant planet.

\subsection{Technique}

Giant-planet frequency can be expressed in terms of metallicity and/or stellar mass with a particular functional
form $f\left(M_{\ast},[Fe/H]\right)$, having a set of free parameters $X$. According to the theorem of Bayes \citep{Bay1763} ,we have

\begin{equation}
P(X|d)P(d) = P(d|X)P(X)
\end{equation}

\noindent where $d$ represents the data. In our case, $d$ can only have two results: a star does have a detectable planet or it does not. For a specific functional form $f$, the best set of parameters to fit the data will be found if $P(X|d)$ is maximized. This means that, given the data, the probability that the parameters $X$ fit the data, is at its maximum.

The probability of the data itself, $P(d)$, is constant. For the parameter probability, $P(X)$, we have chosen to adopt a uniform distribution, allowing us to keep an open mind towards the results. Since these two terms are constant, this means that $P(X|d)$ will be maximized if $P(d|X)$ is maximized.

Now, assume you have $N$ stars in your sample of which $H$ stars are planet hosts. Then you can write $P(d|X)$ as the product of all the separate probabilities per star:

\begin{eqnarray}
P(d|X) & = & \prod_i^N P(d_i|X) \nonumber\\
 & = & \prod_i^H f\left(M_{\ast},[Fe/H]\right) \prod_i^{N-H} \left[1 - f\left(M_{\ast},[Fe/H]\right)\right]. \label{EqProd}
\end{eqnarray}

\noindent If a star has a detectable planet, its probabilty thus equals $f$. In the absence of a giant planet around the star, the probabilty is then obviously $1-f$.

Since the stellar mass $M_{\ast}$ and metallicity $[Fe/H]$ are measured values with errorbars, it is more correct to use a probability
distribution function (pdf) $p(M_i,F_i)$ instead of their actual values. Per star, the giant planet frequency is then actually calculated by

\begin{equation}
f(M_i,F_i) = \int \int p(M_i,F_i)f(M,F)dM dF.
\end{equation}

\noindent The pdf $p$ can be approximated by a product of Gaussians with means $(M_i,F_i)$ and standard deviations $(\sigma_{M_i},\sigma_{F_i})$.

Different functional forms can be evaluated by using the Bayesfactor, as described in \citet{Kass95}. When testing two functional forms $f$ and $g$, the Bayesfactor is expressed as

\begin{equation}\label{EqBay}
B_{fg} = \frac{P(d|f)}{P(d|g)} = \frac{\int P(d|X_f) P(X_f) dX_f}{\int P(d|X_g) P(X_g) dX_g}
\end{equation}

\noindent with $X_f,X_g$ the set of parameters for $f$ and $g$, respectively. The integration limits are the limits of the explored parameter space. According to \citet{Kass95}, a Bayesfactor of about $100$ is needed to statistically rule out a specific functional form.

\subsection{Functional forms}

Previous studies \citep[e.g. ][]{Fis05,Lov07,Udry07,John10,Sou11b} have shown that giant planet frequency is most likely exponentially dependent on metallicity $[Fe/H]$ and polynomially dependent on stellar mass $M_{\ast}$. By including those two dependencies, we adopted the following general functional form:

\begin{equation}\label{pow}
f(M_{\ast},[Fe/H]) = c \left(\frac{M_{\ast}}{M_{\odot}}\right)^{a} 10^{b [Fe/H]}.
\end{equation}

As stated before, there are still questions regarding the correct form in the low-metallicity tail. While it must be noted that there are still many other functional forms that may fit the data, we tested seven different functional forms in our analysis:

\begin{enumerate}
\item The traditional metallicity and mass dependence (Eq. \ref{pow}),
\item Only a metallicity dependence $(a=0)$,
\item Only a mass dependence: $f\cdot M_{\ast}^g + h$,
\item\label{ItAfix} Eq. \ref{pow} with $a$ fixed to $1$,
\item\label{ItPowcd} A combination of several functional forms, like
\begin{equation}\label{powcd}
f(M_{\ast},[Fe/H]) = \left\{ \begin{array}{ll}
c M_{\ast}^{a} 10^{b [Fe/H]} & \mbox{if $[Fe/H] \geq d$} \\
c 10^{bd} & \mbox{if $e \leq [Fe/H] < d$} \\
0 & \mbox{if $[Fe/H] < e$}\qquad,
\end{array}  \right.
\end{equation}
\item\label{ItPowc} Eq. \ref{powcd} with $e=-\infty$,
\item\label{ItPowd} Eq. \ref{powcd} with $d = e$.
\end{enumerate}

Form number \ref{ItAfix} takes both metallicity and stellar mass into account, but fixes the mass dependence to a linear dependence. The last three forms will vary the behavior of the function for the lower metallicities. It is considered that for decreasing metallicity, the function flattens out, resulting in a constant rather than an exponential (parameter $d$). Another consideration is a sudden lack of giant planets. \citet{Me12} suggested that there may be a lower limit in metallicity below which no giant planets can be formed. This lower limit is represented by parameter $e$. Option \ref{ItPowcd} takes both the constant and the stop into account, while options \ref{ItPowc} and \ref{ItPowd} take only the constant, resp. stop into account.

In our parameter space, we assumed a uniform distribution over several intervals as seen in Table \ref{TabPriors}. These assumptions are based on what has been written in the literature in previous studies \citep[e.g. ][]{Lov07,Udry07,John10}, and the derived parameters do not depend on the a priori chosen intervals.

\begin{table}
\caption{Chosen interval for the parameters in the functions.}
\label{TabPriors}     
\centering                       
\begin{tabular}{ccc}
\hline\hline 
Parameter & Interval & Step\\
\hline
$a$ & $[0.0,2.0]$ & 0.1 \\
$b$ & $[0.0,3.0]$ & 0.1 \\
$c$ & $[0.01,0.15]$ & 0.01 \\
$d$ & $[-0.5,0]$ & 0.05 \\
$e$ & $[-0.75,-0.55]$ & 0.05 \\
$f$ & $[0.05,0.55]$ & 0.05 \\
$g$ & $[0.0,2.4]$ & 0.1 \\
$h$ & $[-0.50,-0.05]$ & 0.05 \\
\hline                       
\end{tabular}
\end{table}

\subsection{Results}

Finding the best solution of one functional form, means maximizing the probability that the parameters $X$ fit the data. Since there are many local maxima throughout our parameter space, we chose to adopt the conservative (but time-consuming) method of stepping through a dense grid. The step sizes we used are given in Table \ref{TabPriors}. To transform the products of Equation \ref{EqProd} into sums, we also worked in logarithmic space \citep[see e.g. ][]{John10}.

Comparing the seven different functional forms, leads to interesting results. In Table \ref{TabRes}, the order of functional forms is given for both samples, as well as the combined sample (HARPS + CORALIE). The top functional form is the best, the bottom one the worst. On the right of these functions, the Bayesfactors are represented that compare these two functions. As can be seen, the only functional form that can statistically be ruled out is form number 3, where there is only a mass dependence. The other six forms cannot be distinguished statistically. By multiplying the Bayesfactors, we calculate that the `best' form and the `worst' form of these six function relate with a Bayesfactor of $2.15$, $3.15$, and $4.12$ for the HARPS, the CORALIE, and the combined sample, respectively. According to \citet{Kass95}, these numbers are too low to conclude anything from it.

\begin{table}
\caption{Results of the Bayesian analysis for the three samples. For each sample, the functions are ordered from the best to the worst in the left columns. The Bayesfactors that compare two functions are given in the right columns.}
\label{TabRes}
\centering                       
\begin{tabular}{c|c|c}
\hline\hline 
HARPS & CORALIE & HARPS+CORALIE\\
\hline
Funct. \hspace{1ex}$ B_{fg}$ & Funct. \hspace{1ex}$B_{fg}$ & Funct. \hspace{1ex}$B_{fg}$ \\
\hline
4 $_{\searrow}$  \hspace{2ex} & 4$_{\searrow}$  \hspace{2ex} & 7$_{\searrow}$  \hspace{2ex}\\
\hspace{6ex}  1.02 &  \hspace{6ex} 1.03 &  \hspace{6ex} 1.06 \\
7$^{\nearrow}_{\searrow}$ \hspace{2ex} & 7$^{\nearrow}_{\searrow}$ \hspace{2ex} & 4$^{\nearrow}_{\searrow}$ \hspace{2ex}\\
\hspace{6ex}  1.09 & \hspace{6ex} 1.17 &  \hspace{6ex} 1.20 \\
1$^{\nearrow}_{\searrow}$ \hspace{2ex} & 2$^{\nearrow}_{\searrow}$ \hspace{2ex} & 2$^{\nearrow}_{\searrow}$ \hspace{2ex}\\
\hspace{6ex}  1.00 & \hspace{6ex} 1.01 &  \hspace{6ex} 1.01 \\
2$^{\nearrow}_{\searrow}$ \hspace{2ex}  & 1$^{\nearrow}_{\searrow}$ \hspace{2ex} & 1$^{\nearrow}_{\searrow}$ \hspace{2ex}\\
\hspace{6ex}  1.44 & \hspace{6ex} 1.25 &  \hspace{6ex} 1.19 \\
5$^{\nearrow}_{\searrow}$ \hspace{2ex} & 5$^{\nearrow}_{\searrow}$ \hspace{2ex} & 5$^{\nearrow}_{\searrow}$ \hspace{2ex}\\
\hspace{6ex}   1.34 & \hspace{6ex} 2.07 &  \hspace{6ex} 2.71 \\
6$^{\nearrow}_{\searrow}$ \hspace{2ex} & 6$^{\nearrow}_{\searrow}$ \hspace{2ex} & 6$^{\nearrow}_{\searrow}$ \hspace{2ex}\\
\hspace{6ex}  73.13 & \hspace{6ex} $1.32 \cdot 10^6$ &  \hspace{6ex} $>10^{10}$ \\
3$^{\nearrow}$ \hspace{2ex} & 3$^{\nearrow}$  \hspace{2ex} & 3$^{\nearrow}$  \hspace{2ex}\\
\hline                       
\end{tabular}
\end{table}

\begin{figure}[t!]
\begin{center}
\includegraphics[width=8.9cm]{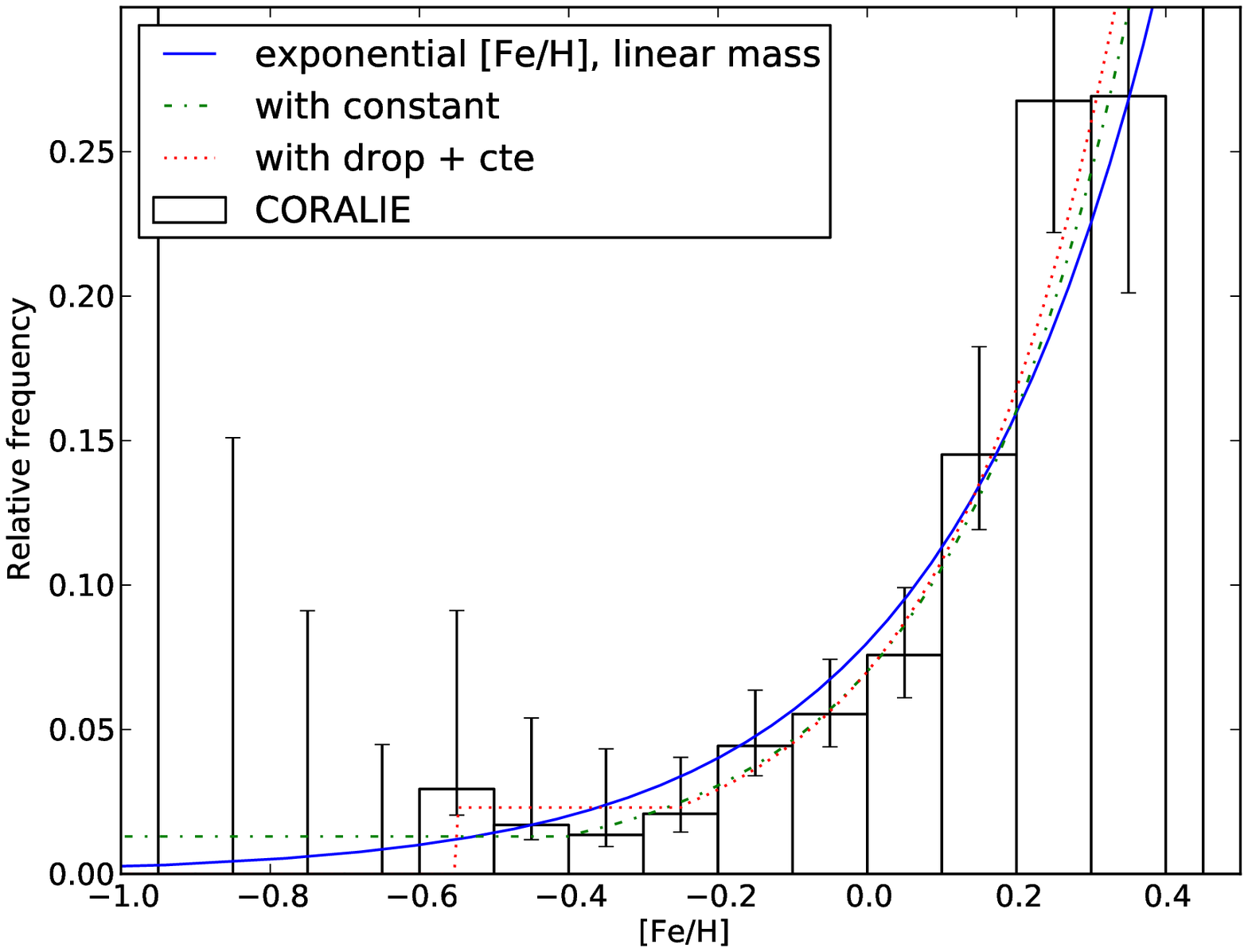}\\
\includegraphics[width=8.9cm]{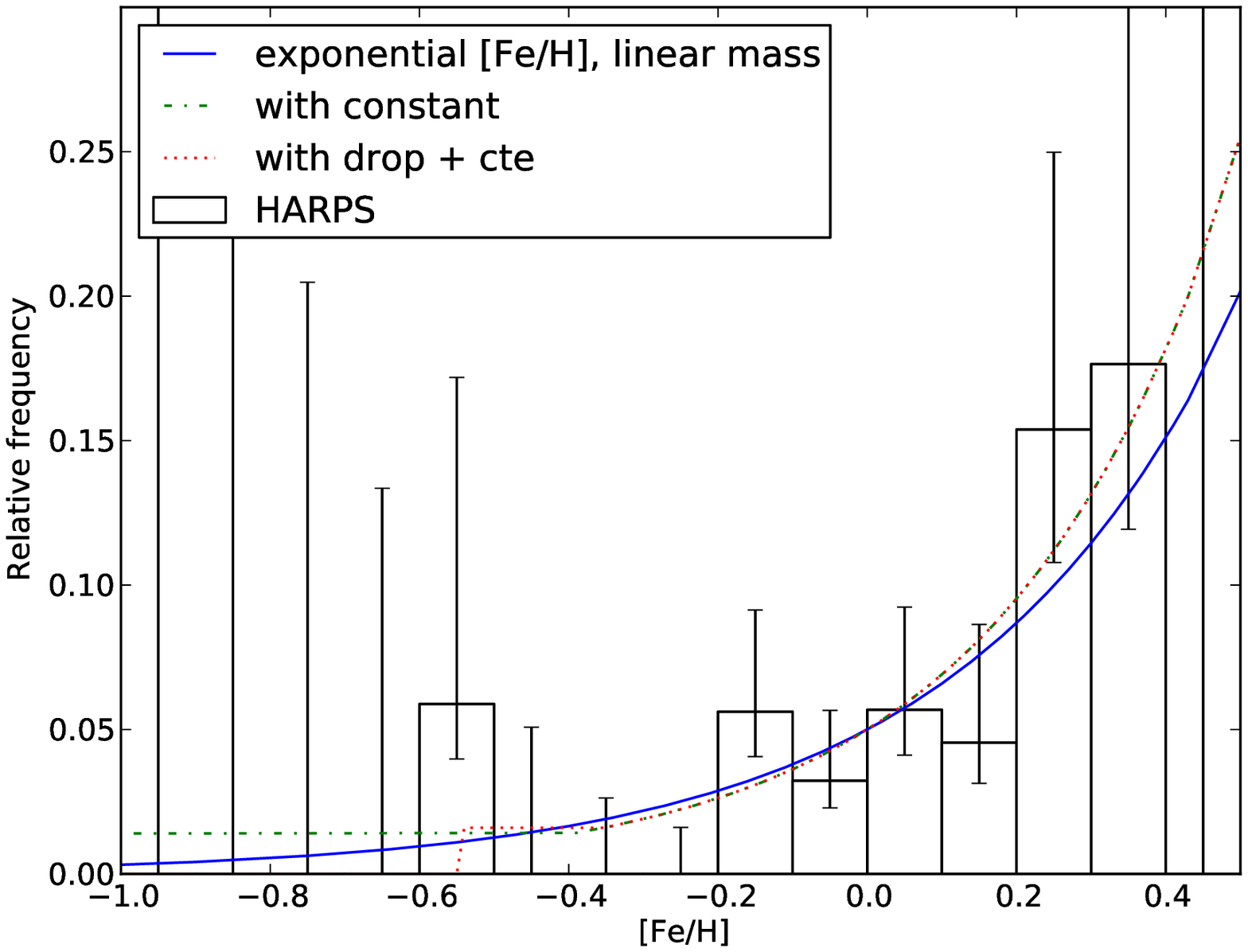}\\
\includegraphics[width=8.9cm]{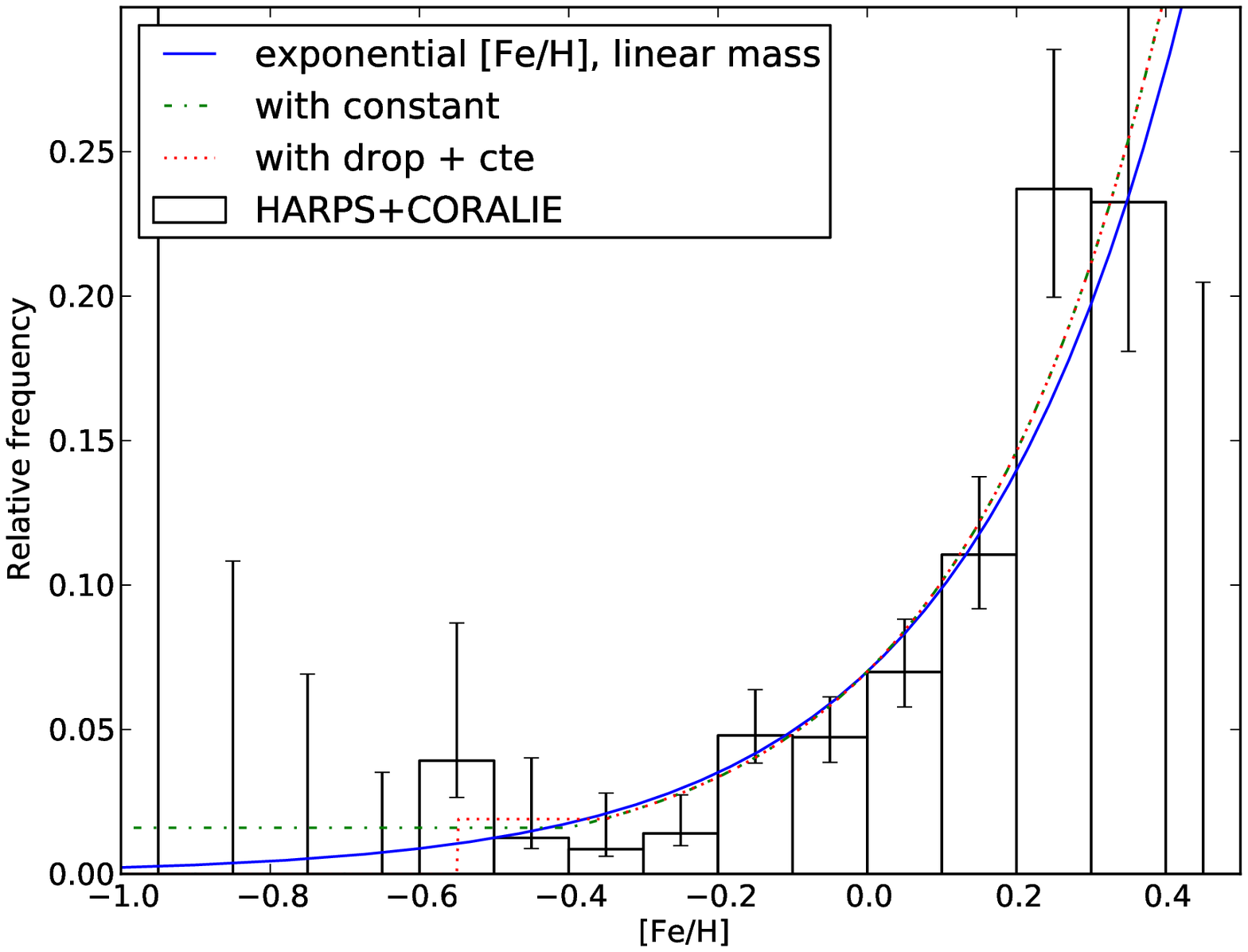}
\caption{Frequency of giant planets as a function of metallicity and mass of the HARPS (top panel), the CORALIE (middle panel) and the combined (bottom panel) sample. Three different functional forms are shown: a complete exponential with linear mass (blue curve), an exponential and a constant (green curve), and an exponential, a constant plus a drop (red curve). The stellar mass is fixed to $M_{\ast} = 1.0 M_{\odot}$.}
\label{FigFreq}
\end{center}
\end{figure}

In Fig. \ref{FigFreq}, the best solutions for three functional forms are shown for both our samples. The functional forms that are shown are the traditional exponential with a linear mass dependence (form nr 4), the exponential with a constant (form nr 6) and the exponential with a constant and a drop (form nr 5). Even from the figures, it is clear that it is hard to distinguish between the different forms in the metal-poor regime.

The conducted planet search surveys \citep{Mayor11} were not complete for all period ranges. This is particularly true for longer periods ($>$1000-2000 days), where the results are incomplete. About a quarter of our sample have these long-period planets. By excluding them, however, we obtain comparable results from the analysis. Our conclusions are thus not sensitive to the inclusions of these planets.

\section{Conclusions and discussion}\label{Con}

We performed a statistical Bayesian analysis to look for the best function to describe giant planet frequency in terms of stellar properties, specifically stellar mass and metallicity. We performed this test with seven different functional forms, based on three volume-limited samples. One sample was observed with CORALIE, the other one with HARPS. The test was also performed for the combined sample.

The test concluded that giant planet frequency is definitely not a function of stellar mass alone. There was no statistically significant result for the other functions. The Bayesfactors, used to compare between the functions, are all between $1.0$ and $4.12$. According to \citet{Kass95}, this means that the difference is not worth more than a bare mention, so it is still unclear what happens exactly in the metal-poor regime.

While the dependence on metallicity is clear, an additional dependence on stellar mass could neither be confirmed nor discarded. The difference between the functional forms was mostly focused on the metal-poor regime. We tested whether the giant planet frequency in that regime is flat rather than exponential. No statistical difference was found between the two options. These results contrast strongly with the results of \citet{John10}. For our analysis, we used the formalism as described by \citet{Kass95} where Bayesfactors are calculated as can be seen in our Eq. \ref{EqBay} and their Eqs. 1-2. Even though \citet{John10} cite the same work, it seems that they use a different formulation for comparing models (Eq. 7 in their work). These different formulas may explain the different results.

It has to be noted, however, that their sample has a much wider mass range, including both A and M stars. This makes their analysis more sensitive to a possible mass dependence. The metallicity range of their sample, however, is comparable to ours.

The very clear trend that giant planet frequency is an increasing function of metallicity in the metal-rich regime can be explained with the theory of core-accretion. The lack of such a trend can be explained more easily with the theory of gravitational instability. In recent works \citep[e.g. ][]{Meru10,Rog12}, simulations have shown that planet formation induced by gravitational instability is more likely to occur around metal-poor stars than around metal-rich ones. 

Disk instability is expected to occur only in the outer regions, explaining among other things the existence of massive long-period planets like HR8799b, c and d \citep{Mar08}. However, this does not exclude that short period planets may also be formed by gravitational instability. \citet{Bar11} suggest that planets formed by gravitational instability may migrate very rapidly inwards. Several giant planets around metal-poor stars have periods between 50 and 1000 days (not very short, not very long). They may also be the result of a gravitationally unstable disk.

These theoretical works suggest that a different trend in giant planet frequency may be expected. If gravitational instability dominates in the metal-poor regime and core-accretion in the metal-rich regime, we would expect to see a constant followed by an exponential. It has to be noted that the samples that we are testing are simply too small to be able to distinguish between a constant or an exponential in the low-metallicity end. We need a much bigger unbiased, volume-limited sample to be able to form any distinctive conclusions. A sample of about 5000 stars will enable us to better distinguish between the different models. With all the current efforts to discover planets in large samples of solar-type stars like GAIA, Kepler, etc., it is likely that this problem will be solved in the near future.

\begin{acknowledgements}
      This work was supported by the European Research Council/European Community under the FP7 through Starting Grant agreement number 239953. NCS also acknowledges the support from Funda\c{c}\~ao para a Ci\^encia e a Tecnologia (FCT) through program Ci\^encia\,2007 funded by FCT/MCTES (Portugal) and POPH/FSE (EC), and in the form of grants reference PTDC/CTE-AST/098528/2008 and PTDC/CTE-AST/09860/2008.
\end{acknowledgements}

\bibliographystyle{aa} 
\bibliography{References.bib}

\end{document}